\documentstyle[11pt]{article}
\begin{document}
\newcommand{\2}{\vspace{0.2 cm}}
\newcommand{\dom}{\mbox{$\rightarrow$}}
\newcommand{\ndom}{\mbox{$\not\rightarrow$}}
\newcommand{\sdom}{\mbox{$\Rightarrow$}}
\newcommand{\nsdom}{\mbox{$\not\Rightarrow$}}
\newcommand{\qed}{\hfill$\Box$}
\newcommand{\pf}{{\bf Proof: }}
\newcommand{\pfs}{{\bf Proof Sketch: }}
\newtheorem{theorem}{Theorem}[section]
\newtheorem{algorithm}[theorem]{Algorithm}
\newtheorem{proposition}[theorem]{Proposition}
\newtheorem{lemma}[theorem]{Lemma}
\newtheorem{problem}[theorem]{Problem}
\newtheorem{corollary}[theorem]{Corollary}
\newtheorem{conjecture}[theorem]{Conjecture}
\newtheorem{remark}[theorem]{Remark}
\newcommand{\beq}{\begin{equation}}
\newcommand{\eeq}{\end{equation}}
\newcommand{\ra}{\rangle}
\newcommand{\la}{\langle}
\newcommand{\har}{\rightleftharpoons}
\newcommand{\<}[1]{\mbox{$\la #1 \ra$}}
\newcommand{\dist}{\mbox{\normalfont dist}}



\title{Parameterized Algorithms for Directed Maximum Leaf Problems\\
(Extended Abstract)}

\date{}

\author{
Noga Alon\thanks{
Department of Mathematics, Tel Aviv University, Tel
Aviv 69978, Israel. E-mail: nogaa@post.tau.ac.il. Research
supported in part by a USA-Israeli BSF grant and by a grant from
the Israel Science Foundation.} \and Fedor V. Fomin\thanks{Department of
Informatics, University of Bergen, POB 7803, 5020 Bergen, Norway.
Email: fedor.fomin@ii.uib.no. Supported by the Norwegian Research Council.} \and Gregory  Gutin\thanks{Department of
Computer Science, Royal Holloway, University of London, Egham,
Surrey, TW20 0EX, UK. Email: gutin@cs.rhul.ac.uk. Research
supported in part by an EPSRC grant.} \and Michael
Krivelevich\thanks{Department of Mathematics, Tel Aviv University,
Tel Aviv 69978, Israel. E-mail: krivelev@post.tau.ac.il.
Research supported in part by a
USA-Israel BSF grant and by a grant from the Israel Science Foundation.} \and Saket
Saurabh\thanks{The Institute of Mathematical Sciences, Chennai,
600 017, India. Email: saket@imsc.res.in} }

\maketitle

\begin{abstract}
We prove that finding a rooted subtree with at least $k$ leaves in a
digraph is a fixed parameter tractable problem. A similar result
holds for finding rooted spanning trees with many leaves in digraphs
from a wide family $\cal L$ that includes all strong and acyclic
digraphs. This settles completely an open question of Fellows and
solves another one for digraphs in $\cal L$. Our algorithms are
based on the following combinatorial result which can be viewed as a
generalization of many results for a `spanning tree with many
leaves' in the undirected case, and which is interesting on its own:
If a digraph $D\in \cal L$ of order $n$ with minimum in-degree at
least $3$ contains a rooted spanning tree, then $D$ contains one
with at least $(n/2)^{1/5}-1$ leaves.
\end{abstract}

\section{Introduction}\label{introsec}
The {\sc Maximum Leaf Spanning Tree} problem (finding
a spanning tree with the maximum number of leaves in a
connected undirected graph) is an intensively studied problem from
an algorithmic as well as a combinatorial point of view
\cite{bonsmaLNCS2747,DingJS01,FellowsMRS00,GalbiatMM97,KleitmanW91,Solis-Oba98}. It fits
into the broader class of spanning tree problems on which hundreds
of papers have been written; see e.g. the  book \cite{BangC03}. It is known to be
NP-hard \cite{GareyJ79}, and APX-hard \cite{GalbiatiMM94}, but can
be fairly well approximated efficiently with multiplicative factor $3$ \cite{LuR98}
and even $2$ \cite{Solis-Oba98}.

In this paper, we initiate the combinatorial and algorithmic study
of two natural generalizations of the problem to digraphs. We say
that a subdigraph $T$ of a digraph $D$ is an {\em out-tree} if $T$
is an oriented tree with only one vertex $s$ of in-degree zero
(called {\em the root}). The vertices of $T$ of out-degree zero are
called {\em leaves}. If $T$ is a spanning out-tree, i.e.
$V(T)=V(D)$, then $T$ is called an {\em out-branching} of $D$. Given
a digraph $D$, the {\sc Directed Maximum Leaf Out-Branching} problem
is the problem of finding in $D$ an {out-branching} with the maximum
possible number of leaves. Denote this maximum by $\ell_s(D)$. When
$D$ has no out-branching, we write $\ell_s(D)=0$. Similarly, the
{\sc Directed Maximum Leaf Out-tree} problem is the problem of
finding in $D$ an {out-tree} with the maximum possible number of
leaves, which we denote by $\ell(D)$. Both these problems are
equivalent for connected undirected graphs, as any maximum leaf tree
can be extended to a maximum leaf spanning tree with the same number
of leaves.

Notice that $\ell(D)\ge \ell_s(D)$ for each digraph $D$. Let $\cal
L$ be the family of digraphs $D$ for which either $\ell_s(D)=0$ or
$\ell_s(D)=\ell(D)$.  It is easy to see that $\cal L$ contains all
strong and acyclic digraphs.

We investigate the above two problems from the \emph{parameterized
complexity} point of view. Parameterized Complexity is a recent
approach to deal with intractable computational problems having some
parameters that can be relatively small with respect to the input
size. This area has been developed extensively during the last
decade. For decision problems with input size $n$, and a parameter
$k$, the goal is to design an algorithm with runtime $f(k)n^{O(1)}$
where $f$ is a function of $k$ alone. Problems having such an
algorithm are said to be fixed parameter tractable (FPT).
The book by Downey and Fellows \cite{downey1999} provides a good introduction to
the topic of parameterized complexity. For recent developments see the books by
Flum and Grohe \cite{FlumGrohebook} and by Niedermeier \cite{Niedermeierbook06}.


The parameterized version of the {\sc Directed  Maximum Leaf
Out-Branching} (the {\sc Directed Maximum Leaf Out-tree}) problem is
defined as follows: Given a digraph $D$ and a
positive integral parameter $k$, is $\ell(D)\geq k$
($\ell_s(D) \geq k$)? We denote the parameterized versions of
the {\sc Directed  Maximum Leaf Out-Branching} and the {\sc Directed Maximum Leaf Out-Tree}
problems by $k$-DMLOB and $k$-DMLOT respectively.


While the parameterized complexity of almost all natural problems on undirected graphs
is well understood, the world of digraphs is still wide open.
The main reason for this anomaly is that most of the techniques developed for undirected graphs
cannot be used or extended to digraphs. One of the most prominent examples is the
{\sc  Feedback Vertex Set} problem, which is easily proved to be FPT for undirected graphs,
while its parameterized complexity on digraphs is a long standing open problem in the area.
In what follows
we briefly explain why the standard techniques for the {\sc Maximum Leaf Spanning Tree}
problem on undirected graphs cannot be used for its generalizations to digraphs.

\begin{itemize}
\item The Graph Minors Theory of Robertson and Seymour
\cite{RobertsonS85a} is a powerful (yet non-constructive)
technique for establishing  membership in FPT. For example,
this machinery can be used to show that the {\sc Maximum Leaf Spanning
Tree} problem is FPT for undirected graphs (see \cite{FellowsL92}). However,
Graph
Minors Theory for digraphs is still in a preliminary stage and
at the moment cannot be used as a tool for tackling interesting directed graph
problems.

\item Bodlaender \cite{Bodlaender89}  used the following
arguments to prove that
the {\sc Maximum Leaf Spanning Tree} problem is  FPT: If an  undirected graph $G$ contains
 a star $K_{1,k}$ as a minor, then it is possible to construct a spanning tree with at
 least $k$ leaves from this minor. Otherwise, there is no
 $K_{1,k}$  minor in $G$, and it is possible to prove  that the
treewidth of $G$ is at most $f(k)$. Thus, dynamic programming
can be used to decide whether there is a tree with $k$ leaves.
This approach does not work  on directed graphs because
containing a big out-tree as a minor does not imply the existence
of an out-branching or out-tree with many leaves in the original
graph.  In short, the properties of having no out-branching  with at least $k$ leaves
or having no out-tree with $k$ leaves are not minor closed.

 \item The seemingly  most efficient approach for
designing FPT algorithms for undirected graphs is based on
a combination of combinatorial bounds and preprocessing rules for
handling vertices of small degrees. Kleitman and West \cite{KleitmanW91} and
Linial and Sturtevant \cite{LinialS87} showed that every connected undirected
graph $G$ on $n$ vertices with minimum degree at least $3$
has a spanning tree with at least $n/4 + 2$ leaves.
Bonsma et al. \cite{bonsmaLNCS2747} combined this combinatorial result
with clever preprocessing rules to obtain the fastest known algorithm
for the $k$-{\sc Maximum Leaf Spanning Tree} problem, running in time
$O(n^3+9.4815^kk^3)$. It is not clear how to devise a similar approach
for digraphs.
\end{itemize}

\medskip
\noindent{\textbf{Our Contribution.}} We obtain a number of
combinatorial and algorithmic results for the {\sc Directed Maximum
Leaf Out-Branching} and  the {\sc Directed Maximum Leaf Out-tree}
problems. Our main combinatorial result (Theorem~\ref{main1}) is the
proof that for every digraph $D\in \cal L$ of order $n$ with minimum
in-degree at least 3, $\ell_s(D)\ge (n/2)^{1/5}-1$ provided
$\ell_s(D)>0$. This can be viewed as a generalization of many
combinatorial results for undirected graphs related to the existence
of spanning trees with many leaves
\cite{GriggsW92,KleitmanW91,LinialS87}.

Our main algorithmic contributions are fixed parameter tractable
algorithms for the $k$-DMLOB and  the $k$-DMLOT problems for
digraphs in $\cal L$ and for all digraphs, respectively. The
algorithms are based on a decomposition theorem which uses ideas
from the proof of the main combinatorial result. More precisely,
we show that either a digraph contains a structure that can be
extended to an out-branching with many leaves, or the pathwidth of
the underlying undirected graph is small. This settles completely
an open question of Mike Fellows \cite{fellows,gutin} and solves
another one for digraphs in $\cal L$.

\section{Preliminaries}

Let $D$ be a digraph. By $V(D)$ and $A(D)$ we represent the vertex
set and arc set of $D$, respectively. An {\em oriented graph} is a
digraph with no directed 2-cycle. Given a subset $V'\subseteq
V(D)$ of a digraph $D$, let $D[V']$ denote the digraph induced on
$V'$. The {\em underlying undirected graph} $UN(D)$ of $D$ is
obtained from $D$ by omitting all orientation of arcs and by
deleting one edge from each resulting pair of parallel edges. The
{\em connectivity components} of $D$ are the subdigraphs of $D$
induced by the vertices of components of $UN(D)$. A vertex $y$ of
$D$ is an {\em in-neighbor} ({\em out-neighbor}) of a vertex $x$
if $yx\in A$ ($xy\in A$, respectively). The {\em in-degree}
$d^-(x)$ ({\em out-degree} $d^+(x)$) of a vertex $x$ is the number
of its in-neighbors (out-neighbors). A vertex $s$ of a digraph $D$
is a {\em source} if the in-degree of $s$ is 0. A strong component
$S$ of a digraph $D$ is a {\em source strong component} if no
vertex of $S$ has an in-neighbor in $V(D)\setminus V(S)$. The
following simple result gives necessary and sufficient conditions
for a digraph to have an out-branching.

\begin{proposition}[\cite{bang2000}]\label{iffoutb}
A digraph $D$ has an out-branching if and only if $D$ has a unique
source strong component.
\end{proposition}

This assertion allows us to check whether $\ell_s(D)>0$ in time
$O(|V(D)|+|A(D)|)$. Thus, we will often assume, in the rest of the
paper, that the digraph $D$ under consideration has an
out-branching.

Let $P=u_1u_2\ldots u_q$ be a directed path in a digraph $D$. An arc
$u_iu_j$ of $D$ is a {\em forward} ({\em backward}) {\em arc for} $P$
if $i\le j-2$ ($j<i$, respectively). Every backward arc of the type $v_{i+1}v_i$
is called {\em double}.

For a natural number $n$, $[n]$ denotes the set $\{1,2,\ldots ,n\}.$

\medskip  The notions of treewidth and pathwidth were introduced by
Robertson and Seymour in  \cite{RobertsonS1} and \cite{RobertsonS2}
(see \cite{Bodlaender89} and \cite{Moehring90} for surveys).

A {\em tree decomposition} of an (undirected) graph $G$ is a pair
$(X,U)$ where $U$ is a tree whose vertices we will call {\em
nodes} and $X=(\{X_{i} \mid i\in V(U)\})$ is a collection of
subsets of $V(G)$ such that
\begin{enumerate}
\item $\bigcup_{i \in V(U)} X_{i} = V(G)$,

\item for each edge $\{v,w\} \in E(G)$, there is an $i\in V(U)$
such that $v,w\in X_{i}$, and

\item for each $v\in V(G)$ the set of nodes $\{ i \mid v \in X_{i}
\}$ forms a subtree of $U$.
\end{enumerate}
The {\em width} of a tree decomposition $(\{ X_{i} \mid i \in V(U)
\}, U)$ equals $\max_{i \in V(U)} \{|X_{i}| - 1\}$. The {\em
treewidth} of a graph $G$ is the minimum width over all tree
decompositions of $G$.

If in the definitions of a tree decomposition and treewidth we
restrict $U$ to be a tree with all vertices of degree at most $2$
(i.e., a path) then we have the definitions of path decomposition
and pathwidth. We use the notation $tw(G)$ and $pw(G)$ to denote the
treewidth and the pathwidth of a graph $G$.

We also need an equivalent definition of pathwidth in terms of
vertex separators with respect to a linear ordering of the vertices.
Let $G$ be a graph and let
$\sigma=(v_1,v_2,\ldots ,v_n)$ be an ordering of $V(G)$. For $j\in
[n]$ put $V_j =\{v_i:\ i\in [j]\}$ and denote by $\partial V_j$
all vertices of $V_j$ that have neighbors in $V\setminus V_j.$
Setting
\[
vs(G,\sigma) = \max_{i\in [n]} |\partial V_i | ,
\]
we define the \emph{vertex separation}  of  $G$ as
\[
vs(G) = \min \{vs(G,\sigma) \colon \sigma \mbox{ is an ordering of
} V(G)\}.
\]

The following assertion is well-known.  It follows directly from
the results of Kirousis and Papadimitriou \cite{KirousisP85} on
interval width  of a graph, see also \cite{Kinnersley92}.

\begin{proposition}[\cite{Kinnersley92,KirousisP85}]\label{sovp_pw_vs}
For any graph $G$, $vs(G)=pw(G)$.
\end{proposition}

\section{Combinatorial Lower Bounds on  $\ell(D)$ and  $\ell_s(D)$}

Let $\cal D$ be a family of digraphs. Notice that if we can show
that $\ell_s(D)\ge g(n)$ for every digraph $D\in {\cal D}$ of order
$n$, where $g(n)$
is tending to infinity as $n$ tends to infinity,
then $k$-DMLOB is FPT on $\cal D$. Indeed,
$g(n)< k$ holds only for digraphs with less than some $G(k)$
vertices and we can generate all out-branchings in such a digraph in
time bounded by a function of $k.$

Unfortunately, bounds of the type $\ell_s(D)\ge g(n)$ are not valid
for all strongly connected digraphs.
Nevertheless, such bounds hold for wide classes of digraphs as we
show in the rest of this section.

The following assertion shows that $\cal L$ includes a large number
digraphs including all strong and acyclic digraphs (and, also,
well-studied classes of semicomplete multipartite digraphs and
quasi-transitive digraphs, see \cite{bang2000} for the definitions).

\begin{proposition}\label{L} Suppose that a digraph $D$ satisfies
the following property: for every pair $R$ and $Q$ of distinct
strong components of $D$, if there is an arc from $R$ to $Q$ then
each vertex of $Q$ has an in-neighbor in $R$. Then $D\in \cal L$.
\end{proposition}
\pf Let $T$ be a maximal out-tree of $D$ with $\ell(D)$ leaves. We
may assume that $\ell_s(D)>0$ and $V(T)\neq V(D).$ Let $H$ be the
unique source strong component of $D$ and let $r$ be the root of
$T.$ Observe that $r\in V(H)$ as otherwise we could extend $T$ by
adding to it an arc $ur$, where $u$ is some vertex outside the
strong component containing $r.$ Let $C$ be a strong component
containing a vertex from $T$. Observe that $V(C)\cap V(T)=V(C)$ as
otherwise we could extend $T$ by appending to it some arc $uv$,
where $u\in V(C)\cap V(T)$ and $v\in V(C)\setminus V(T).$ Similarly,
one can see that $T$ must contain vertices from all strong
components of $D$. Thus, $V(T)=V(D)$, a contradiction.\qed

\subsection{Digraphs with Restricted In-Degree}

\begin{lemma}\label{mlemma}
Let $D$ be an oriented graph of order $n$ with every vertex of
in-degree 2 and let $D$ have an out-branching. If $D$ has no
out-tree with $k$ leaves, then $n\le 2k^5.$
\end{lemma}
\pf Assume that $D$ has no out-tree with $k$ leaves. Consider an
out-branching $T$ of $D$ with $p$ leaves (clearly $p<k$).
As long as the remaining part of $T$ is not a path consider a directed path from the root of
$T$ to a leaf, and omit the part of it that starts right after the last vertex along the
path whose degree in $T$ is at least $3$, and ends at the leaf. This process provides
a collection $\cal P$ of $p$ vertex-disjoint directed paths covering
all vertices
of $D$.

Let $P\in {\cal P}$ have $q\ge n/p$ vertices and let $P'\in {\cal
P}\setminus \{P\}$. There are at most $k-1$ vertices on $P$ with
in-neighbors on $P'$ since otherwise we could choose a set $X$ of at
least $k$ vertices on $P$ for which there were in-neighbors on $P'$.
The vertices of $X$ would be leaves of an out-tree formed by the
vertices $V(P')\cup X.$  Thus, there are $m\le (k-1)(p-1)\le
(k-1)(k-2)$ vertices of $P$ with in-neighbors outside $P$ and at
least $q-(k-2)(k-1)$ vertices of $P$ have both in-neighbors on $P$.

Let $P=u_1u_2\ldots u_q$. Suppose that there are $2(k-1)$ indices
$$i_1<j_1\le i_2<j_2\le \cdots \le i_{k-1}<j_{k-1}$$
such that each
$u_{i_s}u_{j_s}$ is a forward arc for $P$. Then the arcs
$$\{u_{i_s}u_{j_s},u_{j_s}u_{j_s+1},\ldots
,u_{i_{s+1}-1}u_{i_{s+1}}:\ 1\le s\le k-2\}\cup
\{u_{i_{k-1}}u_{j_{k-1}}\}\cup \{u_{i_s}u_{i_s+1}:\ 1\le s\le
k-1\}$$ form an out-tree with $k$ leaves, a contradiction.

Let $f$ be the number of forward arcs for $P$. Consider the graph $G$
whose vertices are all the forward arcs and a pair $u_iu_j,u_su_r$ of
forward arcs are adjacent in $G$ if the intervals $[i,j-1]$ and
$[s,r-1]$ of the real line intersect. Observe that $G$ is an
interval graph and, thus, a perfect graph. By the result of the
previous paragraph, the independence number of $G$ is less than
$k-1$. Thus, the chromatic number of $G$ and the order of its
largest clique $Q$ is at least $f/(k-2)$. Let
$V(Q)=\{u_{i_s}u_{j_s}:\ 1\le s\le g\}$ and let $h=\min \{j_s-1:\
1\le s\le g\}$. Observe that each interval $[i_s,j_s-1]$ contains
$h$. Therefore, we can form an out-tree with vertices
$$\{u_1,u_2,\ldots ,u_h\}\cup \{u_{j_s}:\ 1\le s\le g\}$$ in which
$\{u_{j_s}:\ 1\le s\le g\}$ are leaves. Hence we have ${f \over
k-2}\le k-1$ and, thus, $f\le (k-2)(k-1)$.

Let $uv$ be an arc of $A(D)\setminus A(P)$ such that $v\in V(P).$
There are three possibilities: (i) $u\not\in V(P)$, (ii) $u\in V(P)$
and $uv$ is forward for $P$, (iii) $u\in V(P)$ and $uv$ is backward
for $P$. By the inequalities above for $m$ and $f$, we conclude that
there are at most $2(k-2)(k-1)$ vertices on $P$ which are not
terminal vertices of backward arcs. Consider a path $R=v_0v_1\ldots
v_r$ formed by backward arcs. Observe that the arcs $\{v_iv_{i+1}:\
0\le i\le r-1\}\cup \{v_jv^+_j:\ 1\le j\le r\}$ form an out-tree
with $r$ leaves, where $v^+_j$ is the out-neighbor of $v_j$ on $P.$
Thus, there is no path of backward arcs of length more than $k-1$.

If the in-degree of $u_1$ in $D[V(P)]$ is 2, remove one of the
backward arcs terminating at $u_1$. Observe that now the backward
arcs for $P$ form a vertex-disjoint collection of out-trees with
roots at vertices that are not terminal vertices of backward arcs.
Therefore, the number of the out-trees in the collection is at most
$2(k-2)(k-1)$. Observe that each out-tree in the collection has at
most $k-1$ leaves and thus its arcs can be decomposed into at most
$k-1$ paths, each of length at most $k$. Hence, the original total
number of backward arcs for $P$ is at most $2k(k-2)(k-1)^2+1.$ On
the other hand, it is at least $(q-1)-2(k-2)(k-1)$. Thus,
$(q-1)-2(k-2)(k-1)\le 2k(k-2)(k-1)^2+1.$ Combining this inequality
with $q\ge n/(k-1)$, we conclude that $n\le 2k^5.$\qed

\begin{theorem}\label{main1}
Let $D$ be a digraph in $\cal L$ with $\ell_s(D)>0$.
\begin{enumerate}
\item[(a)] If  $D$ is an oriented graph with minimum in-degree at
least 2, then $\ell_s(D)\ge (n/2)^{1/5}-1.$ \item[(b)] If $D$ is a
digraph with minimum in-degree at least 3, then $\ell_s(D)\ge
(n/2)^{1/5}-1.$
\end{enumerate}
\end{theorem}
\pf (a) Let $T$ be an out-branching of $D$. Delete some arcs from
$A(D)\setminus A(T)$, if needed, such that the in-degree of each
vertex of $D$ becomes 2. Now the inequality $\ell_s(D)\ge
(n/2)^{1/5}-1$ follows from Lemma \ref{mlemma} and the definition of
$\cal L$.

(b) Let $T$ be an out-branching of $D$. Let $P$ be the path formed
in the proof of Lemma \ref{mlemma}. (Note that $A(P)\subseteq
A(T)$.) Delete every double arc of $P$, in case there are any, and
delete some more arcs from $A(D)\setminus A(T)$, if needed, to
ensure that the in-degree of each vertex of $D$ becomes 2. It is not
difficult to see that the proof of Lemma \ref{mlemma} remains valid
for the new digraph $D$. Now the inequality $\ell_s(D)\ge
(n/2)^{1/5}-1$ follows from Lemma \ref{mlemma} and the definition of
$\cal L$.\qed

\2

It is not difficult to give examples showing that the restrictions
on the minimum in-degrees in Theorem~\ref{main1} are optimal.
Indeed, any directed cycle $C$ is a strong oriented graph with all
in-degrees $1$ for which $\ell_s(C)=1$ and any directed double cycle
$D$ is a strong digraph with in-degrees $2$ for which $\ell_s(D)=2$
(a {\em directed double cycle} is a digraph obtained from an
undirected cycle by replacing every edge $xy$ with two arcs $xy$ and
$yx$).

\section{Parameterized Algorithms for $k$-DMLOB and $k$-DMLOT}

In the previous section, we gave lower bounds on $\ell(D)$ and
$\ell_s(D)$ for digraphs $D\in \cal L$ with minimum in-degree at
least $3$. These bounds trivially imply the fixed parameter
tractability of the $k$-DMLOB and the $k$-DMLOT problems for these
class of digraphs. Here we extend these FPT results to digraphs in
$\cal L$ for $k$-DMLOB and to all digraphs for $k$-DMLOT. We prove a
decomposition theorem which either outputs an out-tree with $k$
leaves or provides a path decomposition of the underlying undirected
graph of width $O(k^3)$ in polynomial time.

\begin{theorem}\label{mainth} Let $D$ be a digraph in $\cal L$ with $\ell_s(D)>0.$
Then either $\ell_s(D)\ge k$ or the underlying undirected graph of
$D$ is of pathwidth at most $k^3$.
\end{theorem}
\pf Let $D$ be a digraph in $\cal L$ with $0<\ell_s(D)<k.$ Let us
choose an out-branching $T$ of $D$ with $p$ leaves. As in the
proof of Lemma~\ref{mlemma}, we obtain a collection $\cal P$ of $p
~(<k)$ vertex-disjoint directed paths covering all vertices of
$D$.

For a path $P\in {\cal P}$, let $W(P)$ be the set of vertices not on
$P$ which are out-neighbors of vertices on $P$. If $|W(P)|\geq k$,
then the vertices $P$ and $W(P)$ would form an out-tree with at
least $k$ leaves, which  by the definition of $\cal L$, contradicts
the assumption $\ell_s(D)< k$. Therefore, $|W(P)|<k$. We define
\[
U_1=\{v\in W(P): P\in {\cal P}\}.
\]
Note that
\[
|U_1|\leq p(k-1)\leq (k-1)^2.
\]
Let $D_1$ be the graph obtained from $D$ after applying the
following trimming procedure around all vertices of $U_1$: for every
path $P\in {\cal P}$ and every vertex $v\in U_1\cap V(P)$ we delete
all arcs emanating out of $v$ and directed into $v$ except those of
the path $P$ itself. Thus for every two paths $P,Q\in  {\cal P}$
there is no arc in $D_1$ that goes from $P$ to $Q$.

For $P \in  {\cal P}$ let $D_1[P]$ be the subdigraph of $D_1$
induced by the vertices of $P$.  Observe that $P$ is  a
Hamiltonian directed path in $D_1[P]$. We denote by  $S[P]$ the
set of vertices which are  
heads of forward arcs in $D_1[P]$.

We claim that $|S[P]|\leq (k-2)(k-1)$. Indeed, for each vertex
$v\in S[P]$, delete all forward arcs terminating at $v$ but one.
Observe that the procedure has not changed the number of vertices
which are heads 
 of forward arcs. Also the number of forward
arcs in the new digraph is  $|S[P]|$. As in the proof of
Lemma~\ref{mlemma}, we can show that the number of forward arcs in
the new digraph is at most $ (k-2)(k-1)$.

Define $ U_2=\{v\in S[P]:\ P\in {\cal P}\}. $ Thus
\[
|U_2|\leq p(k-2)(k-1)\leq (k-2) (k-1)^2.
\]
Let $D_2$ be the graph obtained from $D_1$ after applying the
trimming procedure as before around all vertices of $U_2$, that is,
for every path $P\in {\cal P}$ and every vertex $v\in U_2\cap V(P)$
we delete all arcs emanating out of $v$ or directed into $v$
except those of the path $P$.

Put $U=U_1 \cup U_2$. Let $C$ be a connectivity component of
$D_2$. Observe that $C$ consists of a directed path $P=v_1v_2
\dots v_q \in  {\cal P}$ passing through all its vertices, together with its
backward arcs. For every $j\in [q]$  let
 $V_j = \{v_i:\ i\in [j]\}$. If for some $j$ the set $V_j$
contained at least $k$ vertices, say $\{v_1',v_2',\cdots ,v_t'\}$
with $t\geq k$, having in-neighbors in the set $\{v_{j+1},v_{j+2},
\dots, v_q \}$, then $D$ would contain an out-tree with at least $k$
leaves formed by the path $v_{j+1}v_{j+2} \dots v_q$ together with a
backward arc terminating at $v_i'$ from a vertex on the path for
each $1\leq i \leq t$, a contradiction. Thus, for the underlying
undirected graph $UN(C)$ of $C$, $vs(UN(C))\leq k$. By
Proposition~\ref{sovp_pw_vs}, the pathwidth of $UN(C)$ is at most
$k$. Since the pathwidth of a graph is equal to the maximum
pathwidth of its connected components, we have that the pathwidth of
$UN(D_2)$  is at most $k$.

Finally, let $(X_1, X_2, \dots, X_p)$ be a path decomposition of
$UN(D_2)$ of width at most $k$. Then $(X_1\cup U, X_2 \cup U,
\ldots, X_p\cup U)$ is a path decomposition of $UN(D)$ of width at
most $k + |U|\leq k^3$.\qed

\2

\begin{theorem}
\label{impcor} $k$-DMLOB is FPT for digraphs in $\cal L$.
\end{theorem}
\pfs Let $D$ be a digraph in $\cal L$ with $\ell_s(D)>0$. The proof
of Theorem~\ref{mainth} can be easily turned into a polynomial time
algorithm to either build an out-branching of $D$ with at least $k$
leaves or to show that $pw(UN(D))\le k^3$ and provide the
corresponding path decomposition. Now the algorithm follows by a
simple dynamic programming over the decomposition. Alternatively,
the property of containing a directed out-branching with at least
$k$ leaves can be formulated as a monadic second order formula.
Thus, by the fundamental theorem of Courcelle \cite{Courc90,
Courc92}, the  $k$-DMLOB problem for all digraphs $D$ with
$pw(UN(D))\le k^3$ can be solved in $O(f(k) \cdot |V(D)|)$ time,
where $f$ is a function depending only on $k$. \qed

Let $D$ be a digraph and let $R_v$ be the set of vertices reachable
from a vertex $v\in V(D)$ in $D$. Observe that $D$ has an out-tree
with $k$ leaves if and only if there exists a $v\in V(D)$ such that
$D[R_v]$ has an out-tree with $k$ leaves. Notice that each $D[R_v]$
has an out-branching rooted at $v$. Thus, we can prove the following
theorem, using the arguments in the previous proofs.

\begin{theorem}
\label{thm-outtree}
For a digraph $D$ and $v\in V(D)$, let $R_v$ be the set of vertices reachable
from a vertex $v\in V(D)$ in $D$. Then either we have $\ell(D[R_v])\ge k$ or the underlying
undirected graph of $D[R_v]$ is of pathwidth at most $k^3$. Moreover, one can find, in
polynomial time, either an out-tree with at least $k$ leaves in $D[R_v]$, or a path
decomposition of it of width at most $k^3$.
\end{theorem}

To solve $k$-DMLOT, we apply Theorem \ref{thm-outtree} to all the
vertices of $D$ and then either apply dynamic programming over the
decomposition or apply Courcelle's Theorem as in the proof of
Theorem \ref{impcor}. This gives the following:

\begin{theorem}
\label{impcortree}
$k$-DMLOT is FPT for digraphs.
\end{theorem}

We can, in fact, show that the $k$-DMLOB problem for digraphs in
$\cal L$ is linear time solvable for a fixed $k$. To do so, given a
digraph $D\in \cal L$ with $\ell_s(D)>0$ we first apply Bodlaender's
linear time algorithm \cite{Bodlaender96} to check whether the
treewidth of $UN(D)$ is at most $k^3$. If $tw(UN(D))> k^3$ then by
Theorem~\ref{mainth} $D$ has an out-branching with $k$ leaves. Else
$tw(UN(D))\leq k^3$ and we can use Courcelle's Theorem  to check in
linear time whether $D$ has an out-branching of size $k$. This gives
the following:
\begin{theorem}
\label{linear} The $k$-DMLOB problem for digraphs in $\cal L$ is
linear time solvable for every fixed $k$.
\end{theorem}

\section{Concluding Remarks and Open Problems}

We have seen that every digraph $D\in \cal L$ with $\ell_s(D)>0$ of
order $n$ and with minimum in-degree at least $3$ contains an
out-branching with at least $(n/2)^{1/5}-1$ leaves. Combining the
ideas in the proof of this combinatorial result with the fact that
the problem of deciding whether a given digraph in $\cal L$ has an
out-branching with at least $k$ leaves can be solved efficiently for
digraphs of pathwidth at most $k^3$ we have shown that the $k$-DMLOB
problem for digraphs in $\cal L$ as well as the $k$-DMLOT problem
for general digraphs are fixed parameter tractable. The
parameterized complexity of the $k$-DMLOB problem for all digraphs
remains open.

For some subfamilies of $\cal L$, one can obtain better bounds on
$\ell_s(D)$. An example is the class of multipartite tournaments. A
multipartite tournament is an orientation of a complete multipartite
graph. It is proved in \cite{gutinMN40,petrovicDM98} that every
multipartite tournament $D$ with at most one source has an
out-branching $T$ such that the distance from the root of $T$ to any
vertex is at most 4. This implies that $\ell_s(D)\ge {n-1 \over 4}.$
 Also for a
tournament $D$ of order $n$, it is easy to prove that
$\ell_s(D)\ge n-\log_2 n.$ (This bound is essentially tight, i.e.,
we cannot replace the right hand side by $n-\log_2 n +
\Omega(\log_2 \log_2 n)$ as shown by random tournaments; see
\cite{alon2000}, pages 3-4, for more details.)

It seems that the bound $\ell_s(D)\ge (n/2)^{1/5}-1$ is far from
tight. It would be interesting to obtain better bounds for digraphs
$D\in \cal L$ (with $\ell_s(D)>0$) of minimum in-degree at least 3.

\2 \2

\noindent{\bf Acknowledgement.} We thank Bruno Courcelle, Martin
Grohe, Eun Jung Kim and Stephan Kreutzer  for useful discussions of
the paper.


\end{document}